\documentclass[pre,preprint,tightenlines,showpacs,superscriptaddress]{revtex4}

\usepackage{psfrag}
\usepackage{amsmath}
\usepackage{amssymb}
\usepackage{amsfonts}
\usepackage{bm}
\usepackage{graphicx}

\newcommand{\bi}[1]{{\bf #1}}
\newcommand{\rmi}{{\rm i}}
\newcommand{\rmd}{{\rm d}}
\newcommand{\etal}{{\em et al.}}
\newcommand{\Scal}{{\cal S}}
\newcommand{\Ocal}{{\cal O}}
\newcommand{\bwt}{\begin{widetext}}
\newcommand{\ewt}{\end{widetext}}

\begin{document}

\title{Dynamic renormalization group study of a generalized continuum
model of crystalline surfaces}

\author{Rodolfo Cuerno}
\email{cuerno@math.uc3m.es}
\affiliation{Departamento de Matem\'aticas \& GISC, Universidad
Carlos III de Madrid, Avenida Universidad 30, E-28911 Legan\'{e}s, Spain}
\author{Esteban Moro}
\email{moro@thphys.ox.ac.uk}
\affiliation{Theoretical Physics Department,
University of Oxford, 1 Keble Road, OX1 3NP, United Kingdom}

\begin{abstract}
We apply the Nozi\`eres-Gallet dynamic renormalization group (RG)
scheme to a continuum equilibrium model of a $d$-dimensional surface
relaxing by linear surface tension and linear surface diffusion, and
which is subject to a lattice potential favoring discrete values of
the height variable. The model thus interpolates between the
overdamped sine-Gordon model and a related continuum model of
crystalline tensionless surfaces. The RG flow predicts the existence
of an equilibrium roughening transition only for $d=2$ dimensional
surfaces, between a flat low-temperature phase and a rough
high-temperature phase in the Edwards-Wilkinson (EW) universality
class. The surface is always in the flat phase for any other substrate
dimensions $d >2$. For any value of $d$, the linear surface diffusion
mechanism is an irrelevant perturbation of the linear surface tension
mechanism, but may induce long crossovers within which the scaling
properties of the linear molecular-beam epitaxy equation are observed,
thus increasing the value of the sine-Gordon roughening
temperature. This phenomenon originates in the non-linear lattice
potential, and is seen to occur even in the absence of a bare surface
tension term. An important consequence of this is that a crystalline
tensionless surface is asymptotically described at high temperatures
by the EW universality class.
\end{abstract}


\date{\today}

\maketitle

\section{Introduction}
\label{sec:introd}

The dynamics of growing surfaces \cite{alb,krug} has attracted great
interest during the last decade. This is due both to the practical
implications for the control of film quality in thin film production
techniques, and to the fundamental questions it raises in areas of
Physics such as spatially extended systems in the presence of
fluctuations \cite{os}, or scale invariance in non-equilibrium systems
\cite{md}.  It has been observed that a surface growing in the
presence of fluctuations quite generically exhibits time and spatial
scale invariance properties. Specifically, if $h(\bi{r},t)$ denotes
the surface height at time $t$ above a $d$-dimensional substrate
position $\bi{r} \in \mathbb{R}^d$, the height-difference correlation
function $G(r,t) = \langle (h(\bi{r}_0+\bi{r},t)-h(\bi{r}_0,t))^2
\rangle$ grows as $G(r,t) \sim t^{2 \alpha/z}$ for $t^{1/z} \ll r$ and
scales as $G(r,t) \sim r^{2 \alpha}$ for $t^{1/z} \gg r$, where
$\alpha$ and $z$ are referred to as the roughness and dynamic
exponents, respectively. These scale invariance properties imply that
such {\em rough} surfaces share many properties with dynamic critical
phenomena, which allows one to obtain useful information on the former
from studies on the latter, and vice versa.  A particularly
interesting example is provided by the overdamped sine Gordon (sG)
model, which describes the equilibrium fluctuations of a crystal
surface and the features of its roughening transition \cite{sgrev2}:
\begin{equation}
\mu^{-1} \frac{\partial h}{\partial t} = F + \nu \Delta h -
\frac{2\pi V}{a_{\perp}} \sin\left(\frac{2 \pi h}{a_{\perp}} \right)
+ \sqrt{2T \mu^{-1}} \, \eta(\bi{r},t) .
\label{sG}
\end{equation}
In Eq.\ (\ref{sG}), $\Delta$ is the substrate Laplacian, $\mu$ is the
surface mobility, $F$, $\nu$, $V$ and $a_{\perp}$ are positive
constants, $T$ is temperature, and $\eta$ is a zero mean, Gaussian
white noise with correlations $\langle \eta(\bi{r},t) \eta(\bi{r}',t')
\rangle = \delta(\bi{r}-\bi{r}') \delta(t-t')$.  The surface
morphology of a crystal surface described by the sG model thus results
from an interplay between linear surface tension, described by the
term with coefficient $\nu$ in (\ref{sG}), a periodic potential
favoring discrete interface values $h=n \ a_{\perp}$ with $n \in
\mathbb{Z}$, and thermal fluctuations. Moreover, using the same
arguments as in \cite{gl}, for a non-zero value of the homogeneous
driving (flux of aggregating particles) $F$ in (\ref{sG}), an
alternative interpretation of the sG surface is as one minimizing
surface area (for small values of the surface slope), subject to a
flux of aggregating particles such that growth events of the height
take place in integer values. In equilibrium ($F=0$), the sG model is
well known \cite{chw,ng,ngreview,sbm} to feature a roughening
transition when $d=2$: for temperatures above a critical value $T \geq
T_R^{\rm sG}$, the lattice potential is irrelevant and the surface is
rough, in the sense quoted above for the height-difference correlation
function. The corresponding exponent values are those of the
Edwards-Wilkinson (EW) equation \cite{ew}, which is simply the $V = 0$
limit of Eq.\ (\ref{sG}), namely \cite{caveat}
\begin{equation}
\alpha_{\rm EW} = \frac{2-d}{2}, \qquad z_{\rm EW}=2 .
\label{expsew}
\end{equation}
In the case $d=2$ considered, (\ref{expsew}) amounts to $\alpha_{\rm
EW} = 0$, implying $G(r,t) \sim \log r$ for $r \ll t^{1/z}$.  For
temperatures smaller than $T_R^{\rm sG}$, the lattice potential
dominates the large scale properties of the surface setting a finite
correlation length $\xi$, beyond which the height-difference
correlation function $G(r > \xi,t)$ attains a constant value and hence
the surface is flat. The sG roughening transition is of the
Kosterlitz-Thouless class, and hence the {\em continuum} sG model is
related to important {\em discrete} models such as the discrete
Gaussian and the $\cal F$ models \cite{sgrev2,saito}.

As mentioned above, a practical domain which presents many instances
of growing surfaces is the area of thin film production by techniques
such as e.g.\ Molecular-Beam Epitaxy (MBE). In many MBE conditions
\cite{pv}, the main relaxation mechanism on the surface is surface
diffusion, rather than surface tension. In this case, and again
considering the additional effect of a periodic lattice potential, one
is led naturally to the following model of a growing surface
\cite{usprl}, henceforth referred to as the xMBE model
\begin{equation}
\mu^{-1} \frac{\partial h}{\partial t}=F
-\kappa \Delta^2 h - \frac{2\pi V}{a_{\perp}} \sin\left(
\frac{2\pi h}{a_{\perp}} \right)
+ \sqrt{2 T \mu^{-1}} \, \eta(\bi{r},t) .
\label{eq}
\end{equation}
In Eq.\ (\ref{eq}), $\kappa$ is a positive constant, in which case the
corresponding term in the continuum equation does provide ---again for
small surface slopes--- a surface diffusion relaxation mechanism
\cite{mullins}.  All the other terms in Eq.\ (\ref{eq}) have the same
meaning as in (\ref{sG}). Specifically, $F$ is related to the flow of
adatoms onto the substrate. Thus, for $F\neq 0$, Eq.\ (\ref{eq})
provides a non-equilibrium description of a growing surface which
roughens under the effect of thermal fluctuations. In particular, as
shown in \cite{usprl}, the surface described by Eq.\ (\ref{eq})
initially displays RHEED (Reflection High Energy Electron Diffraction)
oscillations, akin
to those observed experimentally \cite{RHEED}. For $F=0$, (\ref{eq})
describes the equilibrium fluctuations at temperature $T$ (we will
consider a unit Boltzmann's constant) of a surface minimizing the
energy functional (Hamiltonian)
\begin{equation}
E(\kappa,V) = \int {\rm d}^d \bi{r} \left\{ \frac{\kappa}{2}
(\Delta h)^2 + V \left[1-\cos\left(\frac{2\pi h}{a_{\perp}}\right)
\right] \right\}.
\label{ham}
\end{equation}
Therefore, such a surface can be also thought of as minimizing surface
curvature to linear approximation, thus having in principle zero
surface tension, and to favor values of the height which are integer
multiples of $a_{\perp}$.  Note in this respect that (\ref{eq}) {\em
does not} have the form of a continuity equation for the surface
height, due to the form of the nonlinear and noise terms
\cite{tn}. Hence, from the point of view of applications to MBE
growth, the continuum xMBE model (\ref{eq}) might be relevant for
those situations in which nonconserved noise is expected to play a
r\^ole, e.g.\ for length-scales larger than the typical diffusion
length \cite{kbkw}.  On the other hand, in a similar way as the sG
equation can be seen as a continuum description of the discrete
Gaussian model, Eq.\ (\ref{ham}) is a natural candidate for the
continuum description of the discrete Laplacian roughening (Lr) model
\cite{nelson} on the square lattice \cite{bruce}. The Lr model has
been used to describe two-dimensional defect melting and as a model of
tensionless surfaces such as membranes
\cite{strand,kleinert}. Actually, Eq.\ (\ref{ham}) can indeed be
obtained \cite{usvar} as a continuum limit of the Lr model, albeit in
a non-unique fashion.

Numerical simulations of Eq.\ (\ref{eq}) for $d=2$ \cite{usprl,tesis}
show an equilibrium roughening transition, similar to that in the sG
model. In the case of Eq.\ (\ref{eq}), the exponents characterizing
the rough high temperature phase are numerically consistent with those
obtained for the $V = 0$ limit of Eq.\ (\ref{eq}) ---the so called
linear MBE ($l$MBE) equation \cite{villain,lds}---, namely
\begin{equation}
\alpha_{l{\rm MBE}} = \frac{4-d}{2}, \qquad z_{l{\rm MBE}} = 4 ,
\label{expslmbe}
\end{equation}
which in $d=2$ implies $\alpha_{l{\rm MBE}}=1$.  However, given that
in the sG system the periodic potential is known to contribute a
correction to the surface tension term upon renormalization, the same
phenomenon is expected to occur in the xMBE model, in which the bare
surface tension is zero. Since surface diffusion is irrelevant in the
presence of surface tension \cite{pv,mank}, this argument together
with the available numerical evidence \cite{usprl,tesis} led to the
prediction that the scaling properties of the xMBE model in the high
temperature phase are the same as those of the sG model.  To date, and
although more recent numerical data \cite{usnum} confirm the existence
of crossover behavior in the critical properties of model (\ref{ham}),
{\em direct} numerical verification of EW scaling properties seems
hard to achieve.  Moreover, previous analytical studies \cite{usvar},
which employed a variational mean-field analysis successfully applied
in \cite{saito2} for the analysis of the sG roughening transition,
have failed to reproduce EW scaling at high temperatures for the xMBE
model.  Specifically, the results obtained in \cite{usvar} include an
upper critical dimension $d_c =4$, below which model
(\ref{eq})-(\ref{ham}) displays a phase transition between a low
temperature flat phase and a high temperature rough phase whose
scaling behavior is that of the $l$MBE equation.  However, the
transition taking place is of {\em first order} for all substrate
dimensions, $d \leq d_c$, while the numerical simulations of
\cite{usprl,tesis} strongly suggest that, at least for $d=2$, the
transition is of a {\em continuous} type.  Besides, the generation of
a surface tension term by the lattice potential referred to above is
due to non-trivial coupling between different surface Fourier modes,
which is unsufficiently accounted for by the variational mean-field
description employed in \cite{usvar}.  Therefore it is natural to try
improve on the mean-field approximation of Ref.\ \cite{usvar} and
explore the scale invariant properties of the system near the phase
transition point through the use of the renormalization group (RG).

In this paper we consider the following generalized continuum model of
surfaces subject to a periodic lattice potential:
\begin{eqnarray}
\mu^{-1} \frac{\partial h}{\partial t} & = & F + \nu \Delta h -
\kappa \Delta^2 h - \frac{2\pi V}{a_{\perp}} \sin\left(
\frac{2\pi h}{a_{\perp}} \right) \nonumber \\
 & & + \sqrt{2 T \mu^{-1}} \, \eta(\bi{r},t) .
\label{eqLangevinGR}
\end{eqnarray}
This equation obviously features both Eqs.\ (\ref{sG}) [sG model] and
(\ref{eq}) [xMBE model] as special cases. Moreover, the linear limit
$V = 0$ \cite{pv,mank} of this equation has been observed to
accurately describe growth experiments of copper aggregates by
electrochemical deposition in the presence of organic additives
\cite{lvb}.  For the sake of simplicity, in this work we will consider
this generalized model in the absence of driving ($F=0$), in which
case Eq.\ (\ref{eqLangevinGR}) describes the equilibrium fluctuations
of a surface with Hamiltonian \bwt
\begin{equation}
E_g(\nu,\kappa,V) = \int {\rm d}^d \bi{r} \left\{
\frac{\nu}{2} (\nabla h)^2 + \frac{\kappa}{2}
(\Delta h)^2 + V \left[1 - \cos\left(\frac{2\pi h}{a_{\perp}}\right)
\right] \right\} .
\label{hamg}
\end{equation}
\ewt In order to study the critical properties of the xMBE model in
$d$ substrate dimensions, we will extend the {\em dynamic} RG approach
devised by Nozi\`eres and Gallet (NG) for the $d=2$ sG model \cite{ng}
(see also a detailed account in \cite{ngreview}) to the generalized
system (\ref{eqLangevinGR}). There are two reasons for our pursuing
this approach: {\em (i)} as anticipated above, in the RG analysis of
(\ref{eq})-(\ref{ham}) a finite (non-zero) surface tension term needs
to be allowed for, given that it is generated in any perturbative
scheme even if its bare amplitude is zero; {\em (ii)} a {\em static}
renormalization group study of the equilibrium system (\ref{ham}) is
ill-defined in some parameter ranges due to divergent integrals
\cite{tesis}, similarly to the sG case \cite{ng}. Still, the static RG
study will provide us, via the appropriate generalization, with the
correct expansion of the model non-linearity in terms of relevant
operators through the use of Kadanoff's operator product expansion
(OPE) \cite{kadanoff}, as was accomplished in \cite{kdo,ng} for the sG
model.  In any case, the results to be obtained from the dynamic RG
study that follows will also cover the case of a system minimizing
{\em both} surface area and surface curvature, and will in particular
allow us to analyze how the standard sG roughening transition is
modified when an additional surface diffusion term is considered.  We
will finally consider the renormalization properties of the xMBE
model, which corresponds to a specific choice of bare parameters
within this generalized framework, and will compare the conclusions
obtained with those from both the variational approach \cite{usvar}
and numerical simulations \cite{usprl,tesis}.  To our knowledge, ours
is the first RG approach to the xMBE model as formulated by
(\ref{eq})-(\ref{ham}), and it may contribute to the elucidation of
the existence and nature of the phase transition in this and related
systems, as the Lr model. There also exist static and dynamic RG
studies of similar systems.  Specifically, the equilibrium properties
of a model which is different from (\ref{eq})-(\ref{ham}) but is
believed to provide the continuum description of the Lr model on the
triangular lattice, have been analyzed in \cite{LD}, and its dynamical
properties have been obtained in \cite{KP} and references
therein. Within the rough MBE surfaces context, the dynamic properties
of the conserved sG model have been studied both under conserved
\cite{tn,smgg} and non-conserved \cite{tn,rk} noise. Finally, a
Langevin equation believed to describe a restricted curvature model
\cite{kds} has been analyzed in \cite{ch} by using RG techniques.

This paper is organized as follows. In Sec.\ \ref{sec:dynrg} we apply
the dynamic RG scheme of NG to Eq.\ (\ref{eqLangevinGR}). The
parameter flow thus obtained for this generalized model is studied in
Sec.\ \ref{sec:gm} as a function of the substrate dimension $d$. The
special limit of Eq.\ (\ref{eqLangevinGR}) corresponding to
crystalline tensionless surfaces, Eq.\ (\ref{eq}), requires additional
considerations of a technical nature, and is deferred to Sec.\
\ref{sec:xMBE}.  Finally, Sec.\ \ref{sec:disconcl} is devoted to
further discussion of the results obtained in the previous sections,
and to summarizing our conclusions.  We additionally provide two
appendices. In Appendix \ref{app:ope} we detail, following
\cite{kdo,ng}, the OPE which is needed in the dynamic RG in order to
perform the appropriate expansion of the lattice potential in
(\ref{eqLangevinGR}) into relevant operators. Appendix
\ref{app:sGtrans} closes with a discussion of the specific way in
which the roughening transition of the sG model generalizes into that
to be obtained in Sec.\ \ref{sec:gm} for
(\ref{eqLangevinGR})-(\ref{hamg}).

\section{Dynamic renormalization group analysis}
\label{sec:dynrg}

This section is devoted to the analysis of Eq.\ (\ref{eqLangevinGR})
in the equilibrium case $F=0$, employing the dynamic RG scheme of NG
\cite{ng,ngreview}.  In this scheme, a coarse graining procedure is
performed over the microscopic modes of the noise term. Namely, the
noise is split into two statistically independent parts,
$\eta=\bar{\eta} + \delta \eta$, such that the total noise power
spectrum is the sum of the corresponding contributions.  Here $\delta
\eta(\bi{r}) \equiv \int_{\bar \Lambda}^{\Lambda} {\rm d}^2 \bi{k}\
{\rm e}^{\rmi \bi{k}\bi{r}} \hat \eta(\bi{k})$, where $\Lambda$ is a
momentum cut-off related e.g.\ to atomic positions on the substrate,
$\bar \Lambda = {\rm e}^{-\varepsilon} \Lambda$ with $\varepsilon$ a
small parameter, and $\hat{\eta}(\bi{k})$ is the spatial Fourier
transform of the noise. Then, microscopic fluctuations are integrated
out by defining $\bar{h} \equiv \langle h(\bar{\eta}+\delta \eta)
\rangle_{\delta \eta}$ and $\delta h \equiv h - \bar{h}$, and by
seeking an equation of motion for the thus defined long distance modes
$\bar{h}$.  The result will be an equation with the same shape as
(\ref{eqLangevinGR}), but with new (renormalized) coefficients, which
are sensitive to the microscopic fluctuations $\delta \eta$ by the
action of the nonlinearities.  Specifically, the dynamic equations for
the $\bar{h}$ and $\delta h$ modes read
\begin{eqnarray}
\mu^{-1} \frac{\partial \bar{h}}{\partial t} & = & \nu \Delta \bar{h} -
\kappa \Delta^2 \bar{h} - \langle \Phi(\bar{h},\delta h) \rangle_{\delta \eta}
\nonumber \\
 & & + \sqrt{2 D} \, \bar \eta(\bi{r},t) , \label{2eqsrg1} \\
\mu^{-1} \frac{\partial \delta h}{\partial t}&= & \nu \Delta \delta h -
\kappa \Delta^2 \delta h - [\Phi(\bar{h},\delta h) - \langle \Phi(\bar{h},
\delta h) \rangle_{\delta \eta}] \nonumber \\
 & & + \sqrt{2 D} \, \delta \eta(\bi{r},t) ,
\label{2eqsrg2}
\end{eqnarray}
where we have defined $D \equiv T \mu^{-1}$, and we have introduced
$\Phi(\bar{h},\delta h) \equiv (2\pi V/a_{\perp}) \sin [2\pi
(\bar{h}+\delta{h})/a_{\perp}]$.  In order for (\ref{2eqsrg1}) to be a
closed equation in $\bar h$, we need to solve for $\delta h$ in
(\ref{2eqsrg2}) and introduce the result into (\ref{2eqsrg1}).  The
formal solution of (\ref{2eqsrg2}) reads
\begin{eqnarray}
\delta h(\bi{r},t) & = & \int \rmd^d\bi{r}' \int_{-\infty}^{t} \rmd t'
\chi_0(\bi{r}-\bi{r}',t-t') \nonumber \\
 & & \times [\sqrt{2D} \, \delta \eta(\bi{r}',t') + \Phi'-\langle
\Phi' \rangle_{\delta \eta}] ,
\label{soleqrg1}
\end{eqnarray}
where the primed notation denotes dependence on the $\bi{r}'$, $t'$ variables,
and the $d$-dimensional free propagator reads
\begin{equation}
\chi_0(\bi{r},t) = \int \frac{\rmd^d k \ \rmd \omega}{(2\pi)^{d+1}}
\frac{{\rm e}^{\rmi (\bi{k}\bi{r} - \omega t)}}{\nu k^2 +
\kappa k^4 - \rmi \omega \mu^{-1}} .
\label{freeprop}
\end{equation}
Due to the non-linear lattice potential, an explicit solution of
(\ref{2eqsrg2}) can only
be obtained by performing a perturbative expansion
in powers of $V$. Thus, defining
\begin{equation}
\delta h(\bi{r},t) = \delta h^{(0)}(\bi{r},t) + V \delta h^{(1)}(\bi{r},t)
+ {\cal O}(V^2) ,
\label{expdeltah}
\end{equation}
we obtain
\begin{eqnarray}
\delta h^{(0)}(\bi{r},t)&= &
\int \rmd^d \bi{r}' \int_{-\infty}^t \rmd t'\
\chi_0(\bi{r}-\bi{r}',t-t') \nonumber \\
 & & \times \sqrt{2D} \, \delta \eta(\bi{r}',t') \label{soldeltah1} , \\
\delta h^{(1)}(\bi{r},t)&= & - \frac{4\pi^2}{a_{\perp}^2}
\int \rmd^d\bi{r}' \int_{-\infty}^t \rmd t'\
\chi_0(\bi{r}-\bi{r}',t-t') \nonumber \\
 & & \times \delta {h'}^{(0)}
\ \cos \left(\frac{2\pi \bar{h}'}{a_{\perp}}\right) \label{soldeltah2} .
\end{eqnarray}
Using $\sin(a+b) = \sin a \cos b + \cos a \sin b$, and within our
perturbation expansion, we can now evaluate in (\ref{2eqsrg1})
\begin{eqnarray}
\langle \Phi(\bar h,\delta h) \rangle_{\delta \eta} & = & \frac{2\pi V}{a_{\perp}}
\left[1- \frac{2\pi^2}{a_{\perp}^2} \langle (\delta h^{(0)})^2
\rangle_{\delta \eta} \right. \nonumber \\
 & & - \left. \frac{4\pi^2 V}{a_{\perp}^2} \langle \delta h^{(0)}
\delta h^{(1)} \rangle_{\delta \eta} + \Ocal (V^2) \right] \nonumber \\
& & \times \sin \left(\frac{2\pi\bar{h}}{a_{\perp}} \right) , \label{phiexpansion}
\end{eqnarray}
which has the form
\begin{equation}
\langle \Phi \rangle_{\delta \eta}(\bar h) = V \Phi^{(1)}(\bar h) +
V^2 \Phi^{(2)}(\bar h) + {\cal O}(V^3) ,
\label{phiexp2}
\end{equation}
with
\begin{eqnarray}
\Phi^{(1)}(\bar h)&=& \displaystyle{\frac{2\pi}{a_{\perp}}
\left[1- \frac{(2\pi)^{2-d} {\cal S}_d \ T \ \Lambda^{d-2} \varepsilon}
{2 a_{\perp}^2 (\nu+\kappa \Lambda^2)} \right]} \nonumber \\
 & & \times \displaystyle{\sin \left(\frac{2\pi\bar{h}}{a_{\perp}} \right)} ,
\label{coefphiexp2a} \\
\Phi^{(2)}(\bar h)&=& \displaystyle{-\frac{8\pi^3}{a_{\perp}^3} \
\langle \delta h^{(0)} \delta h^{(1)} \rangle_{\delta \eta} \;
\sin \left(\frac{2\pi\bar{h}}{a_{\perp}} \right)},
\label{coefphiexp2b}
\end{eqnarray}
where $\Scal_d = 2\pi^{d/2}/\Gamma(d/2)$ is the surface area of the
unit hypersphere in $d$ dimensions, $\Gamma(\cdot)$ being Euler's
Gamma function.  The shape of (\ref{coefphiexp2a}) already reflects
the fact that we have considered an infinitesimal shell of microscopic
noise modes of width $\varepsilon$. We can write (\ref{coefphiexp2b})
more explicitly as
\begin{eqnarray}
\Phi^{(2)}(\bar h) & = & \frac{32\pi^5}{a_{\perp}^5} \int \rmd^d\bi{r}'
\int_{-\infty}^{t} \chi_0(\bi{r}-\bi{r}',t-t') \label{dh0dh1} \\
 & & \times \langle \delta h^{(0)} \delta {h'}^{(0)} \rangle_{\delta \eta}
\sin \left(\frac{2\pi \bar h}{a_{\perp}} \right) \
\cos \left(\frac{2\pi\bar h'}{a_{\perp}} \right) .
\nonumber
\end{eqnarray}
Neglecting higher order harmonics of the lattice potential,
$\sin\left(2\pi\bar{h}/a_{\perp}\right) \ \cos
(2\pi\bar{h}'/a_{\perp}) \simeq \frac{1}{2} \sin[2\pi(\bar{h}-
\bar{h}')/a_{\perp}]$. Further, we can use the results for the OPE of
the lattice potential obtained in Appendix \ref{app:ope}, and a Taylor
expansion to obtain
\begin{eqnarray}
\Phi^{(2)}(\bar h)&\simeq&\frac{32\pi^6}{a_{\perp}^6} \Scal_d
\int_0^\infty \rho^{d-1}\ \rmd \rho \int_0^\infty \rmd \tau\:
\chi_0(\rho,\tau) \label{expsin} \\
 & & \times
\left\{\tau \frac{\partial \bar h}{\partial t} -\frac{\rho^2}{2d} \Delta \bar h
- \frac{\rho^4}{8d(d+2)}\Delta^2 \bar h\right\} \nonumber \\
 & & \times \langle \delta h^{(0)} \delta {h'}^{(0)} \rangle_{\delta \eta}
\exp\left[-\frac{2\pi T}{a_{\perp}} \phi(\rho,\tau,\nu,\kappa,\mu)
\right] , \nonumber
\end{eqnarray}
where we have defined $\bm{\rho} \equiv \bi{r} - \bi{r}'$ and $\tau
\equiv t - t'$, the $d$-dimensional free propagator reads
\begin{equation}
\chi_0(\rho,\tau)=\frac{\mu}{(2\pi)^{d/2}}\int_0^\Lambda \rmd k \,
\frac{k^{d/2}}{\rho^{d/2-1}} \, J_{d/2-1}(\rho k) \, {\rm e}^{-(\nu k^2 +
\kappa k^4)\mu \tau} ,
\label{freeprop2}
\end{equation}
and we have introduced
\begin{eqnarray}
\phi(\rho,\tau,\nu,\kappa,\mu) & \equiv & \frac{\Scal_d}{(2\pi)^{d-1}}
\int_{0}^{\Lambda} \rmd k \, \frac{k^{d-1}}{\nu k^2 + \kappa k^4} \bigg[1 -
\nonumber \\
 & - &
\frac{\Gamma(d/2)J_{d/2-1}(\rho k)}{(\rho k/2)^{d/2-1}}\ {\rm e}^{-(\nu k^2 +
\kappa k^4)\mu \tau}\bigg],
\label{phidinam}
\end{eqnarray}
with $J_n(x)$ being the $n$-th order Bessel function of the first
kind.  Finally, using the results of Eqs.\ (\ref{coefphiexp2a}) and
(\ref{expsin}), Eq.\ (\ref{phiexp2}) has the form
\begin{eqnarray}
\langle \Phi \rangle_{\delta h}(\bar h) & = & (V + \varepsilon \, \delta V)
\sin \left(\frac{2\pi\bar h}{a_{\perp}}\right)
- \varepsilon \, \delta \mu^{-1} \, \frac{\partial \bar h}{\partial t}
\nonumber \\
 & & + \varepsilon \, \delta \nu \, \Delta \bar h + \varepsilon \, \delta \kappa
\, \Delta^2 \bar h + \Ocal(V^3) , \label{exp_epsilon}
\end{eqnarray}
where the corrections $\delta \nu$, $\delta \kappa$, etc.\ are implicitly
defined. Inserting the result of Eq.\ (\ref{exp_epsilon}) into
Eq.\ (\ref{2eqsrg1}), we obtain that, to $V^2$ order, the long
distance modes $\bar h$ obey an equation with the same shape as Eq.\
(\ref{eqLangevinGR}), namely,
\begin{equation}
\tilde{\mu}^{-1} \frac{\partial \bar{h}}{\partial t} =
\tilde{\nu} \, \Delta \bar{h} -
\tilde{\kappa} \, \Delta^2\bar{h} - \frac{2\pi \tilde{V}}{a_{\perp}}
\sin \left(\frac{2\pi\bar{h}}{a_{\perp}}\right) + \sqrt{2 D} \,
\bar \eta(\bi{r},t) ,
\label{new1eq}
\end{equation}
but with {\em new coefficients} $\tilde{\nu} \equiv \nu + \varepsilon
\, \delta \nu$, $\tilde{\kappa} \equiv \kappa + \varepsilon \,
\delta\kappa$, $\tilde{\mu}^{-1} \equiv \mu^{-1} + \varepsilon \,
\delta \mu^{-1}$, and $\tilde{V} \equiv V + \varepsilon \, \delta V$.
In order to recover the original Fourier mode cut-off $\Lambda$, we
now rescale variables as
\begin{eqnarray}
\bi{r}& \to & \bi{r}'= \bi{r}/b \\
\bar h & \to & \bar h'= \bar h b^{-\alpha} \\
t & \to & t'= b^{-z} t
\end{eqnarray}
where $b = \mathrm{e}^{\varepsilon}$, and we thus get
\begin{equation}
\bar \mu^{-1} \frac{\partial \bar{h}}{\partial t} =
\bar \nu \, \Delta \bar{h} - \bar \kappa \,
\Delta^2\bar{h} - \frac{2\pi \bar V}{a_{\perp}} \sin
\left(\frac{2\pi\bar{h}}{a_{\perp}}\right) + \sqrt{2
\bar D}\:\bar \eta(\bi{r},t) ,
\label{new2eq}
\end{equation}
with coefficients $\bar \nu =
\tilde \nu \, b^{z-2}$, $\bar \kappa = \tilde \kappa \,
b^{z-4}$, $\bar V = \tilde V \,  b^{z-2\alpha}$,
$\bar \mu^{-1}= \tilde \mu^{-1}$, and
$\bar D =  D \, b^{z-2\alpha-d}$. Finally, for an infinitesimal
$\varepsilon = \mathrm{d} l$ and expanding $\bar \nu(\varepsilon)$, etc., to
first order in $\varepsilon$, we obtain the dynamic RG flow to $V^2$ order:
\begin{eqnarray}
\frac{\rmd \nu}{\rmd l}& = & (z-2) \nu \label{RGdinam1} \\
 & & + \frac{(2\pi)^{6-d}\, \Scal_d\, T\,
\Lambda^{d-6}}{4 \, d \, a_{\perp}^6 (\nu + \kappa \Lambda^2)^2}
\, V^2\, B^{(2,0)}(\nu,\kappa) ,
\nonumber \\
\frac{\rmd \kappa}{\rmd l}&= & (z-4)\kappa \label{RGdinam2} \\
 & & - \frac{(2\pi)^{6-d}\, \Scal_d\, T\, \Lambda^{d-8}}
{16 d(d+2) a_{\perp}^6 (\nu + \kappa \Lambda^2)^2}
\, V^2\, B^{(4,0)}(\nu,\kappa) , \nonumber \\
\frac{\rmd V}{\rmd l}&= & (z-2\alpha)V -
\frac{(2\pi)^{2-d}\, \Scal_d\, T\, \Lambda^{d-2}}
{2 a_{\perp}^2(\nu+\kappa \Lambda^2)}\, V ,
\label{RGdinam3} \\
\frac{\rmd \mu^{-1}}{\rmd l} & = & \frac{(2\pi)^{6-d}\, \Scal_d\, T\,
\Lambda^{d-6}} {2 a_{\perp}^6 (\nu+\kappa \Lambda^2)^3}\: V^2\,
\mu^{-1} \, B^{(0,1)}(\nu,\kappa) ,
\label{RGdinam4} \\
\frac{\rmd D}{\rmd l}&= & (z-2\alpha-d) D , \label{RGdinam5} \\
\frac{\rmd a_{\perp}}{\rmd l}&= & -\alpha a_{\perp} ,
\label{RGdinam6}
\end{eqnarray}
with
\bwt
\begin{eqnarray}
B^{(n,m)}(\nu,\kappa) & \equiv & \int_0^{\infty} \rmd\tilde{\rho}
\, \tilde{\rho}^{n+1}\int_0^{\infty} \rmd\tau
\, {\rm e}^{-\tau} \,  \tau^{m} \,
J_{d/2-1}(\tilde{\rho}) \:
G(\tilde{\rho},\tau,\nu,\kappa) \, \exp \left[-\frac{2\pi T}{a_{\perp}^2}
\phi(\tilde \rho,\tau,\nu,\kappa,\mu) \right] ,
\label{gdinam} \\
G(\tilde{\rho},\tau,\nu,\kappa)& \equiv & \int_0^{1}  \rmd\tilde{k} \,
\tilde{k}^{d/2} \, J_{d/2-1}(\tilde{\rho} \tilde{k})
\exp\left[-\tau\frac{\nu\tilde{k}^2+
\kappa\Lambda^2\tilde{k}^4}{\nu + \kappa \Lambda^2}\right] ,
\end{eqnarray}
\ewt where $\tilde{\rho} \equiv \rho \Lambda$. It is worth noting that
the flow of the mobility $\mu$ is enslaved to that of all other system
parameters, which allows us to neglect its evolution under
(\ref{RGdinam1})-(\ref{RGdinam6}) in the rest of the paper. Moreover,
in order to preserve fluctuation-dissipation by this RG procedure
\cite{ng,ngreview} we need to impose the exponent relation $z = 2
\alpha + d$ in the flow (\ref{RGdinam1})-(\ref{RGdinam6}). Note that
this relation (termed {\em hyperscaling} \cite{wv} in the studies of
kinetic roughening) holds exactly at two fixed lines of
(\ref{RGdinam1})-(\ref{RGdinam6}), which are the two linear limits of
(\ref{eqLangevinGR}), namely the EW [$V=\kappa=0$, see Eq.\
(\ref{expsew})] and $l$MBE [$V = \nu = 0$, see Eq.\ (\ref{expslmbe})]
equations. Both systems can be interpreted as describing equilibrium
fluctuations, governed by the corresponding Hamiltonians. We also
remark that, as anticipated in Sec.\ \ref{sec:introd}, it is clear
from Eq.\ (\ref{RGdinam1}) that, if we consider the RG flow for the
xMBE model (\ref{eq})-(\ref{ham}) in which there is no bare surface
tension, the lattice potential does generate it under RG iterarion.

Finally, we find it more convenient to express the flow in the dimensionless
variables
\begin{equation}
x \equiv \frac{2 a_{\perp}^2}{\pi T}(\nu + \kappa \Lambda^2),\quad
y \equiv \frac{4\pi V}{T\Lambda^2}, \quad K \equiv \frac{\kappa a_{\perp}^2
\Lambda^2}{\pi T}.
\end{equation}
Thus, we have
\begin{eqnarray}
\frac{\rmd x}{\rmd l}&= & (d-2) x -4 K \label{finalflujodin1} \\
 & & + \frac{4}{d}\, \frac{y^2}{x^2}\, \left[\tilde B^{(2)}(x,K)
-\frac{1}{4(d+2)} \tilde B^{(4)}(x,K)\right] , \nonumber \\
\label{finalflujodin2}
\frac{\rmd y}{\rmd l}&= &  2y \left[\frac{d}{2}-\frac{\Scal_d}{(2\pi)^{d-1}}
\frac{\Lambda^{d-2}}{x}\right], \\
\label{finalflujodin3}
\frac{\rmd K}{\rmd l}&= & (d-4)K - \frac{1}{2 d (d+2)}\,  \frac{y^2}{x^2}
\, \tilde B^{(4)}(x,K) ,
\end{eqnarray}
where
\bwt
\begin{eqnarray}
\tilde{B}^{(n)}(x,K) & \equiv & \frac{\Scal_d\, \Lambda^{d-2}}{(2\pi)^{d-1}}
\int_0^{\infty} \rmd\tilde{\rho}\: \tilde{\rho}^{n+1}
\int_{0}^{\infty} \rmd\tau\, {\rm e}^{-\tau} J_{d/2-1}(\tilde{\rho}) \,
\tilde{G}(\tilde{\rho},\tau,x,K) \,  {\rm e}^{-2 \tilde{\phi}(\tilde{\rho},
\tau,x,K)} , \label{Btilde} \\
\tilde{G}(\tilde{\rho},\tau,x,K)&\equiv & \int_{0}^1 \rmd\tilde{k}\,
\tilde{k}^{d/2}\,
J_{d/2-1}(\tilde{k}\tilde{\rho}) \: {\rm e}^{-2\tau
[(x/2-K)\tilde{k}^2+K \tilde{k}^4 ]/x} , \label{Gtilde} \\
\tilde{\phi}(\tilde{\rho},\tau,x,K)& \equiv &\frac{\Scal_d \, \Lambda^{d-2}}
{(2\pi)^{d-1}} \int_{0}^1 \rmd\tilde{k} \,
\frac{\tilde k^{d-1}}{(x/2-K)\tilde{k}^2+K\tilde{k}^4}
\left[1-\frac{\Gamma(d/2) \, J_{d/2-1}(\tilde k \tilde \rho)}{(\tilde k
\tilde \rho/2)^{d/2-1}}\
{\rm e}^{-2\tau
[(x/2-K)\tilde{k}^2+K \tilde{k}^4 ]/x}\right] .
\label{phitilde}
\end{eqnarray}
\ewt
Note that the convergence properties of the integral $\tilde
\phi(\tilde \rho,\tau,x,K)$ defined above depend on the substrate
dimensionality $d$ and on the parameter values. Specifically, for
$d=2$ {\em and} the condition associated with the xMBE model (\ref{eq}),
namely, $x = 2K$, the integral diverges logarithmically at the lower limit.
This divergence originates in the fact that correlation functions do not
have a well-defined thermodynamic limit for the $l$MBE model in $d=2$ \cite{pv}.

The dynamic RG flow (\ref{finalflujodin1})-(\ref{finalflujodin3}) just
obtained generalizes that in \cite{ng,ngreview} for the case of a
finite surface diffusion term, and for any value of the substrate
dimension $d$.  The sine-Gordon case is retrieved simply by setting
$\kappa = 0$ and neglecting the higher order contribution $\tilde
B^{(4)}(x,0)$, related with the RG flow of the surface diffusion term.
Actually, this integral will allow us to study in the next section the
effect of such relaxation mechanism on the sG roughening transition.

\section{Discussion of the RG flow} \label{sec:disc}

The fixed point structure of the RG flow
(\ref{finalflujodin1})-(\ref{finalflujodin3}) depends on substrate
dimensionality $d$, and thus so do the critical properties of model
(\ref{eqLangevinGR})-(\ref{hamg}). In what follows we will restrict
ourselves to dimensions $d \geq 2$ in which phase transitions are
expected to occur.

For general values of $d$, there are no proper non-trivial
fixed points of (\ref{finalflujodin1})-(\ref{finalflujodin3}),
but rather regions in the $(x,y,K)$ phase space which are
invariant under the RG flow. These correspond to the two significant
linear limits of the generalized model, namely the EW line, $\Gamma_{\rm EW}$,
and the $l$MBE line, $\Gamma_{l{\rm MBE}}$, defined by
\begin{eqnarray*}
\Gamma_{\rm EW} & = & \{ (x,y,K): y=K=0, x \neq 0 \} , \nonumber \\
\Gamma_{l{\rm MBE}} & = & \{ (x,y,K): x=2K, y = 0 \} .
\end{eqnarray*}
However, for the special dimensions $d=2$ and $d=4$,
respectively, these regions become lines of fixed points.

\subsection{Generalized model} \label{sec:gm}

We start by analizing generic features of the RG flow as a function of
substrate dimension $d$, in the case in which the bare values of
surface diffusion and surface tension are {\em both} non-zero. The
special initial condition (for the RG flow) in which there
is no bare surface tension (xMBE model) will be considered in detail
in Sec.\ \ref{sec:xMBE}.

\subsubsection{$d=2$ case}

Obviously, the case which is most interesting from the physical point
of view is that of a two-dimensional substrate. As stated above, now
the invariant EW region $\Gamma_{\rm EW}$ actually becomes the
well-known sG line of fixed points \cite{sgrev2,ngreview}. An
important point along this line is $x=1$, where the flow of the
lattice potential $y$ changes stability, see Appendix
\ref{app:sGtrans}. If the bare value of the surface diffusion
$\kappa(l=0) = 0$, as in the sG model, numerical integration of
(\ref{finalflujodin1})-(\ref{finalflujodin3}) shows that the flow
essentially remains on the $(x,y)$ plane and thus completely reduces
to that of the sG system, as seen in panel (a) on Fig.\ \ref{fig:2}:
for high $T$ the flow is towards the $y=0$ axis, the interface being
rough and characterized by the scaling properties of the EW
equation. For low $T$ values, the lattice potential $y$ grows
indefinitely upon iteration of the RG flow, drawing the system into
the $T=0$ limit for which the interface is flat. Thus, there exists a
roughening transition between these flat and rough
phases. Numerically, we estimate the critical temperature as the value
characterizing the trajectory that separates between flow onto the
$y=0$ line and flow towards increasing $y$.  We thus obtain $T_R^{sG}
\simeq 0.725\pm 0.05$ for $V = 1, a_{\perp} = 1, \nu = 1, \kappa=0$,
which is close to the value for the flow equations of the pure sG
model \cite{sgrev2,ng}, $T_R^{sG} = 2/\pi \simeq 0.64$.  On the other
hand, as we see in panels (b) and (c) on Fig.\ \ref{fig:2}, in general
a small initial surface diffusion shifts the roughening temperature to
higher values. The reason for this effect is that, as studied in
\cite{pv,mank} when $V = 0$, even though surface diffusion is
irrelevant relative to surface tension in the hydrodynamic limit, an
initial value $\kappa(0) \neq 0$ introduces a crossover length scale,
$L_{\times}(l) = [\kappa(l)/\nu(l)]^{1/2}$, below which surface
diffusion is the relevant relaxation mechanism. Once the
coarse-graining procedure has overcome this scale, $\kappa(l)$
renormalizes to zero (see Appendix \ref{app:sGtrans}) and the flow
takes place on the sG $(x,y)$ plane with a $\nu$ coefficient which
differs from its bare value. From there on the flow is effectively as
that of the pure sG model. The results shown on panel (d) of Fig.\
\ref{fig:2} correspond to a numerical simulation of the sG model [Eq.\
(\ref{eqLangevinGR}) with $\kappa \equiv 0, F = 0$] for $\mu=\nu = V =
a_{\perp} = 1$, and of Eq.\ (\ref{eqLangevinGR}) for $\kappa=0.5$, for
both of which we have computed the specific heat [defined as $\chi_E
\equiv (\langle E_g^2 \rangle - \langle E_g \rangle^2)/(T L^2)$, with
$E_g$ as in (\ref{hamg})] for several substrate lateral sizes $L$ and
periodic boundary conditions.  Clearly, the roughening transition
temperature ---which is preceded by a peak in $\chi_E$ as in the
pure sG model \cite{swendsen}--- shifts to
higher values for $\kappa \neq 0$. Integrating numerically the RG flow
(\ref{finalflujodin1})-(\ref{finalflujodin3}), we obtain
$T_R^{sG}(\kappa=0.5) \simeq 1.35 \pm 0.05$.

\subsubsection{$d > 2$ cases}

In order to inquire about the upper critical dimension, $d_c$, of
model (\ref{eqLangevinGR})-(\ref{hamg}), we consider values of $d$
above the physical two-dimensional case. An important consequence of
this is that the points on the EW line $\Gamma_{\rm EW}$ are not fixed
any longer under the RG iteration. This already signals the final
result that, in these dimensions, there is no proper phase transition.
Rather, the only existing phase is the flat low temperature one.
Before justifying this result in detail, let us note that the EW
equation already predicts a flat surface for $d >2$
\cite{alb,krug,caveat}, and therefore when adding a lattice potential
to the surface tension term in the sG equation, the result that
$d_c=2$ is the corresponding upper critical dimension
\cite{kadanoff,uppsG} does not come as a surprise.  In the generalized
model (\ref{eqLangevinGR})-(\ref{hamg}), similar results to those in
the previous section indicate that the same value $d_c=2$ actually
occurs: for $2 < d < 4$, it is clear from Eq.\ (\ref{finalflujodin3})
that $K(l)$ still decreases exponentially under the RG iterarion, and
again the flow becomes essentially that of the sG model in the
corresponding dimension, thus predicting a flat morphology. In Fig.\
\ref{fig:2new}(a) we show as an illustration results of a numerical
integration of (\ref{finalflujodin1})-(\ref{finalflujodin3}) for
$d=3$. As is clear from the figure [for this, it is useful to note the
projection of the RG flow lines onto the $(x,y)$ plane], the flow is
eventually towards large $y$ values for all temperatures. Note also
that the $K(l)$ value may in some cases become negative, signalling a
non-physical instability.  We attribute this effect [also apparent on
panels (b) and (c) of Fig.\ \ref{fig:2new}] to limitations of our
$\Ocal(V^2)$ approximation to the RG flow.

The value $d=4$ is marginal for the surface diffusion term in Eq.\
(\ref{eqLangevinGR}). In an analogous way to the r\^ole of $d=2$ for
the EW equation [see Eq.\ (\ref{expsew})], the $l$MBE equation
predicts a roughness exponent $\alpha=0$ for $d=4$ [logarithmic
behavior of $G(r,t)$, see Eq.\ (\ref{expslmbe})] and a flat morphology
for $d>4$ \cite{alb,krug}.  Even though the decay of $K(l)$ under the
RG flow (\ref{finalflujodin1})-(\ref{finalflujodin3}) might not be so
fast as in smaller dimensions, it nevertheless occurs, see Fig.\
\ref{fig:2new}(b), with the result of an effective sG behavior in
$d=4$ dimensions, again corresponding to a flat morphology. Finally,
as illustrated in Fig.\ \ref{fig:2new}(c) for $d=5$, if $d>4$ all
coefficients in the RG flow grow under iteration.  In particular the
lattice potential $y$ increases indefinitely and thus the surface
morphology is dictated by the behavior of the $T \to 0$ limit, namely
again there is no phase transition and the surface is flat.

\subsection{Crystalline tensionless surfaces} \label{sec:xMBE}

As remarked above, within the RG flow of the generalized model
(\ref{eqLangevinGR})-(\ref{hamg}), the case corresponding to the xMBE
model (\ref{eq})-(\ref{ham}) simply amounts to a specific condition on
the bare parameters, namely they lie on the plane $x(0)= 2 K(0)$,
$y(0) \neq 0$, thus implying $\nu(0)=0$.
In this section we thus study the RG flow for the xMBE model separately
for different values of the substrate dimension.

\subsubsection{$d=2$ case} \label{sec:xMBEd=2}

In the physical $d=2$ case, and due to the divergencies in the
correlation functions mentioned above, this very condition induces
a trivial RG flow, since the $\tilde B^{(n)}$ integrals in
(\ref{Btilde}) become identically zero. The linear combination
$x-2K$ ---which is proportional to the surface tension---, is a
constant under the RG flow. Equations
(\ref{finalflujodin1})-(\ref{finalflujodin3}) can be exactly
solved to show that any initial condition on the $x=2K$ plane
tends to the origin for $l \rightarrow \infty$.  For large enough
but finite $l$, the surface scales as the $l$MBE equation {\em
independently} of the value of $T$, and thus there is no
temperature driven phase transition, which contradicts the results
of e.g.\ numerical simulations of Eq.\ (\ref{eq}) for
two-dimensional substrates \cite{usprl}.  Thus, some kind of
integral regularization is needed when $d=2$. Here we introduce a
lower momentum cut-off $1/(\Lambda L)$ in the integrals
(\ref{Gtilde}), (\ref{phitilde}), with $L$ being a measure of the
lateral dimension of the two-dimensional substrate. Proceeding in
this way, it can be seen that the integrals (\ref{Btilde}) now
vanish as a power law of $1/L$. In this regularization scheme, the
RG iteration has to be stopped once ${\rm e}^l = L$. Another
possible procedure like that of dimensional regularization can be
employed with the same conclusions as those that follow, similar
to the RG analysis of tensionless membranes \cite{piran}.

In any case, the regularized integrals $\tilde{B}^{(n)}(x,x/2)$ are no
longer identically zero. This leads us to expect the RG flow to escape
from the $x=2K$ plane, which would mean a finite surface tension has
been generated under renormalization, and thus the occurence of {\em
temperature dependent} behavior.  Nevertheless, the flow may take many
iterations before it appreciably deviates from the $x=2K$ plane, given
that, in its neighborhood, the integrals $\tilde{B}^{(n)}$ can be
rather small numbers for large $L$ values.

In Fig.\ \ref{fig:1} we show the numerical integration of the RG flow
(\ref{finalflujodin1})-(\ref{finalflujodin3}) for various initial
conditions on the xMBE plane and $L=128$.  Indeed, three different
types of behavior can be distinguished.  For high enough values of
$T$, the variables $x$, $y$, $K$ are very small and the RG flow
escapes from the $x=2K$ plane very slowly. Under these conditions, the
system flows towards the origin while featuring the scale invariant
behavior of the $l$MBE line $\Gamma_{l{\rm MBE}}$.
An example is trajectory (1) on Fig.\ \ref{fig:1}.  For intermediate
temperatures, $\nu \sim x - 2K$ becomes non-negligible and the flow is
attracted by the $(x,0,0)$ segment with $x<1$, where the behavior is
described by the EW equation, see trajectory (2) on the
figure. Finally, for high values of $T$, the flow falls rapidly onto
the $(x,y)$ plane with $x>1$, see trajectory (3) on the figure.  This
behavior is described by the low temperature massive phase of the sG
model, where the lattice potential is dominant and the surface is
flat, see Appendix \ref{app:sGtrans}.

In order to examine this behavior more closely, we can focus on the
crossover length $L_{\times}$ introduced in Sec.\ \ref{sec:gm}.
By coarse-graining the system with the RG transformation for
scales up to the system size ($l^* = \ln L$), and if the lattice
potential turns out to become irrelevant, we can decide whether the
system scaling is of the $l$MBE or EW type by evaluating
$L_{\times}(l^*)$ to be larger or smaller than $L(l^*) \equiv 1$,
respectively. Note that, for a zero lattice potential, the RG flow of
$L_{\times}$ is easily computed from (\ref{RGdinam1})-(\ref{RGdinam2})
to be given by $L_{\times}(l) = L_{\times}(0) \exp(-l)$, hence an
exponential behavior of $L_{\times}(l)$ will be a sign of the
irrelevance of the lattice potential at scale $l$. In Fig.\
\ref{fig1'} we show the RG flow of $L_{\times}(l)$ for the same
temperature conditions as in Fig.\ \ref{fig:1} and $L(l=0) = 128$.  As
we see, for high enough temperatures [lines (1), (2) in the figure],
indeed $y(l)$ is seen to decay very rapidly to zero, $L_{\times}(l)$
displaying an exponential behavior. However, while for the highest
temperature condition [line (1)] the scaling behavior is of the $l$MBE
type since $L_{\times}(l^*) > 1$, we observe that there are lower
temperatures for which the lattice potential is irrelevant, but the
long distance behavior is rather of the EW type, since
$L_{\times}(l^*) < 1$. This is the case of e.g.\ the RG flow line (2)
on Figs.\ \ref{fig:1} and \ref{fig1'}, for which the system falls onto
the high temperature line of fixed points of the sG model. Finally,
for low enough temperatures the lattice potential dominates the
asymptotic properties of the system, as signalled in Fig.\ \ref{fig1'}
by the complex non-exponential behavior of $L_{\times}(l)$ for
$T=0.45$.  These temperature condition corresponds in Fig.\
\ref{fig1'} to the RG flow of line (3), along which the system falls
onto the flat low temperature phase of the sG model.  In analogy with
the sG case, the dynamic RG flow
(\ref{finalflujodin1})-(\ref{finalflujodin3}) thus predicts a
roughening transition at a temperature
\begin{equation}
T_R = \frac{2a_{\bot}^2 (\nu+\kappa \Lambda^2)}{\pi} ,
\label{Ttrans}
\end{equation}
at which the flow of the lattice potential changes stability.  Note
both $\nu$ and $\kappa$ in (\ref{Ttrans}) are renormalized, rather
than bare values.  For $T < T_R$, the system is in the sG massive low
temperature phase, within which the surface is flat. For $T \geq T_R$,
the surface is rough and should asymptotically feature the EW scale
invariant behavior. However, crossover behavior exists; specifically,
for a small system size (for which the asymptotic behavior may not be
reached) or for very high temperatures [for which the RG flow needs
many iterations in order to escape the xMBE plane of initial
conditions, see line (1) in Fig.\ \ref{fig:1}], the scaling behavior
observed in the system will be of the $l$MBE type.

\subsubsection{$d > 2$ cases}

We have also considered the RG flow of the xMBE model for higher
substrate dimensions, finding a behavior which is qualitatively
consistent with all the above analysis. Moreover, given that for any
$d>2$ the integral $\tilde \phi$ in (\ref{phitilde}) is always a
finite number, the results that follow can be taken as additional
support for those presented in $d=2$ employing lattice cut-off
regularization. As can be seen in Fig.\ \ref{fig:5new}, for $d>2$ the
RG flow of the xMBE model does take on the exact shape of the
generalized model studied in Sec.\ \ref{sec:gm}, and all the
conclusions drawn there become applicable also for the specific
initial condition we are now considering. Namely, the RG flow escapes
from the $x=2K$ plane, and thereon the behavior is similar to that of
the sG model. In particular, there is no phase transition and the
surface morphology is flat for any value $d>2$. Therefore, the upper
critical dimension of the xMBE model (\ref{eq})-(\ref{ham}) is also
predicted to be $d_c =2$. Specifically, in Fig.\ \ref{fig:5new}, we
plot results of a numerical integration of
(\ref{finalflujodin1})-(\ref{finalflujodin3}) starting from the xMBE
model initial condition for substrate dimensions $d=3$ [panel (a)] and
$d=4$ [panel (b)].  As can be seen in the figure, for any initial
condition on the xMBE plane ---that is, for any temperature value---,
the RG flow is towards the sG plane and towards increasing $y$ values,
which is the behavior that corresponds to the low temperature flat
phase, as in the sG model for the corresponding substrate
dimension. For a given value of $d$, the higher the temperature is,
the longer will it take for the RG flow to escape from the plane of
initial conditions, since the crossover length $L_{\times}$ is
larger. In this very high temperature conditions, the initial $l$MBE
behavior will be relevant for a longer time in the dynamics of Eq.\
(\ref{eq}).

\section{Summary and Conclusions} \label{sec:disconcl}

The dynamic RG study presented for the generalized model of
crystalline surfaces (\ref{eqLangevinGR})-(\ref{hamg}) predicts an
upper critical dimension $d_c=2$. Thus, for a two-dimensional surface
there exists a roughening temperature, in such a way that the high
temperature phase exhibits the scaling properties of the EW
equation. Moreover, the transition properties are controlled by the sG
fixed point (see Appendix \ref{app:sGtrans}), and is thus expected to
be of a continuous type.  If we think of
(\ref{eqLangevinGR})-(\ref{hamg}) as a generalization of the sG model
in which a (possibly small) surface diffusion term has been allowed
for, we have seen in Sec.\ \ref{sec:gm} that the consequence is an
increase in the sG roughening temperature, due to the crossover
induced by such an irrelevant perturbation. Moreover, the specific
case of the xMBE model has been seen to share all these properties,
its peculiarity of having a zero bare surface tension merely
introducing much more severe crossover effects. From the point of view
of applications to epitaxial growth systems, this result illustrates
the relevance of EW scaling as an universality class in MBE: in
principle, if the symmetry of the system prohibits non-equilibrium
surface currents and includes invariance of the dynamics under
arbitrary surface tilts, $l$MBE scaling is expected \cite{krug}.
However, we have obtained for a system relaxing linearly as in the
$l$MBE equation that, if a lattice potential influences the dynamics
as in (\ref{eq}) ---thereby accounting e.g.\ for the discrete
character of deposition events or the influence of an underlying
lattice---, then asymptotic EW scaling should occur. Admittedly,
crossovers associated with $l$MBE behavior may be nonetheless rather
long, particularly for high temperatures.

For the case of the xMBE model, the existence of a phase transition
between a flat low temperature phase and a rough high temperature
phase as predicted by (\ref{finalflujodin1})-(\ref{finalflujodin3}) is
compatible with previous results obtained by the variational
approximation \cite{usvar} and numerical simulations
\cite{usprl,tesis}.  However, while the variational mean field
predicts the transition to be first order, the incorporation of
fluctuations by means of the RG is consistent with a phase transition
of a continuous type, which is closer to the numerical
results. Moreover, our present RG study predicts the upper critical
dimension to be that of the sG model, $d_c=2$, while mean-field
predicts $d_c=4$. We note that the absence of non-trivial mode
coupling in the variational mean-field approach limits its
capabilities in determining correctly both lower and upper critical
dimensions, see e.g.\ \cite{saito2} and \cite{usvar} for the case of
the sG model.  Concerning numerical simulations \cite{usprl,tesis}, EW
scaling has been obtained in the high temperature phase only for
temperatures which are extremely close to the roughening temperature,
whereas $l$MBE behavior has been obtained for all other temperature
values above the transition.  Recall we obtained in Sec.\
\ref{sec:xMBEd=2} that, if the temperature was high enough, the system
might need a long time to overcome the crossover associated with
$l$MBE behavior.  Seemingly, the finite system sizes thus far employed
in the simulations ($L \lesssim 128$) are affected by this type of
crossover limitations.  We also note that the occurrence of EW scaling
for the xMBE model is reminiscent of the hexatic phase claimed for the
Lr model \cite{nelson,bruce,strand}.  This is an intermediate phase
which features EW scaling, and which lies in between the flat low
temperature phase (the liquid phase in the melting context) and the
high temperature phase (solid phase) with $l$MBE scaling, being
separated from both of them by roughening transitions of the KT type
\cite{nelson,bruce,strand}.  On the other hand, as found in
\cite{janke} and references therein, there are also claims on the
inexistence of an hexatic phase in the Lr model and, moreover, on the
first order type of the only roughening transition ensuing.  We are
currently performing large scale simulations of the Lr model on the
square lattice \cite{usnum} in order to check the predictions of our
RG analysis, in particular whether any trace of crossover behavior and
EW scaling can be detected in the surface properties at high
temperatures.

\begin{acknowledgments}
We are grateful to Juan Jes\'us Ruiz-Lorenzo and Angel S\'anchez
for discussions and encouragement, and to Martin Rost for discussions
and comments. E.M.\ aknowledges the EU fellowship
No.\ HPMF-CT-2000-0487. This research has been supported by EPSRC (UK)
grant No.\ GR/M04426, DGES (Spain) grant No.\ HB1999-0018, and
by MCyT (Spain) grant No.\ BFM2000-0006.
\end{acknowledgments}

\appendix

\section{}
\label{app:ope}

In this Appendix we recall the considerations needed to perform the
correct Taylor expansions in (\ref{dh0dh1}), using Kadanoff's operator
algebra \cite{kadanoff}. We generalize results for the static RG
analysis of the sine-Gordon nonlinearity \cite{kdo,ng}. Thus, in our
dynamical RG calculation each of the (infinite number of)
non-linearities $O_{2n+1}(\bm{\rho}) =
(\bar{h}-\bar{h}')^{2n+1}$ appearing in the Taylor expansion of $\sin
[2 \pi(\bar{h}- \bar{h}')/a_{\perp}]$ contributes a term proportional
to the marginal operator $O_{1}(\bm{\rho}) =
\bar{h}-\bar{h}'$, where, as above, $\bm{\rho} = \bi{r}-\bi{r}'$.
In our continuum approach, this is the most
relevant term originating (via Taylor expansion) the renormalization
of {\em both} the $\Delta h$ and $\Delta^2 h$ terms in Eq.\
(\ref{eqLangevinGR}). Thus,
\begin{eqnarray}
O_{2n+1}(\bm{\rho}) & \equiv &
(\bar{h}(\bi{r})-\bar{h}(\bi{r}'))^{2n+1} \label{exporper} \\
 & = & \tilde{O}_{2n+1}(\bi{r}) + a_{2n+1}(\bm{\rho})(\bar{h}-\bar{h}') ,
\nonumber
\end{eqnarray}
where $\tilde{O}_{2n+1}(\bi{r})$ is an irrelevant operator. A way to
compute the $a_{2n+1}$ constants is \cite{kdo,kadanoff} by performing
all contractions contributing to the behavior
\begin{equation}
\langle O_{2n+1}(\bi{r}) O_1(\bi{r}+\bm{\rho}) \rangle_{\bar{h}} \sim
a_{2n+1}(\bm{\rho}) \langle O_{1}(\bi{r}) O_1(\bi{r}+\bm{\rho}) \rangle_{\bar{h}} .
\nonumber
\end{equation}
where, within our order of approximation in powers of $V$, averages
$\langle \cdots \rangle_{\bar h}$ are computed with respect to the
Gaussian  distribution of $\bar h$ given by the $V=0$ limit of
(\ref{hamg}).  Thus one obtains
\begin{equation}
a_{2n+1}(\bm{\rho}) = (2n+1) \langle (\bar{h}-\bar{h}')^{2n} \rangle_{\bar{h}}
= \frac{(2n+1)!}{2^n n!} \langle (\bar{h}-\bar{h}')^{2}
\rangle^n_{\bar{h}}
\nonumber
\end{equation}
Using this result in the Taylor expansion of the sine, one gets
\bwt
\begin{eqnarray}
\sin \left[\frac{2\pi(\bar{h}-\bar{h}')}{a_{\perp}}\right]& =&
\sum_{n=0}^{\infty} \frac{(-1)^n}{(2n+1)!} \
\left(\frac{2\pi}{a_{\perp}}\right)^{2n+1} \! \!
O_{2n+1}(\bm{\rho}) \nonumber \\
& \sim & \sum_{n=0}^{\infty} \frac{(-1)^n}{2^n n!} \
\left(\frac{2\pi}{a_{\perp}}\right)^{2n+1}
\! \! \langle (\bar{h}-\bar{h}')^{2}
\rangle^n_{\bar{h}} \  O_1(\bi{r}) \nonumber \\
& = & \frac{2\pi}{a_{\perp}} \, O_1(\bi{r}) \ \exp
\left[-\frac{2\pi^2}{a_{\perp}^2} \langle (\bar{h}-\bar{h}')^2
\rangle_{\bar{h}}\right] .
\label{senope}
\end{eqnarray}
\ewt
This is the result employed in Eq.\ (\ref{expsin}) of Sec.\ \ref{sec:dynrg}.

\section{}
\label{app:sGtrans}

As already stated, the dynamic RG flow
(\ref{finalflujodin1})-(\ref{finalflujodin3}) can be
seen as a generalization of that for the sG equation, as derived e.g.\ in
\cite{ng,ngreview}. In this Appendix we explore this relationship in
some more detail in the physically interesting case $d=2$.

Similarly to the sG case, an important point in the $(x,y,K)$ parameter
space is that where the lattice potential $y$ changes stability. Actually, the
existence in the generalized model of a surface diffusion term does not change
this fact, in the sense that the point $(1,0,0)$ still controls the
behavior of the RG flow to a large extent. This is due to the facts that
the surface diffusion is irrelevant with respect to surface tension, and that
the latter is generated by the RG flow,
as obtained in Sect.\ \ref{sec:dynrg}. The simplest way to
substantiate this conclusion is to study the flow
(\ref{finalflujodin1})-(\ref{finalflujodin3})
perturbatively near the point $(1,0,0)$. To this end, we introduce the
temperature-like variable \cite{ng,ngreview} $t \equiv 2/x -2$, rewrite
Eqs.\ (\ref{finalflujodin1})-(\ref{finalflujodin3})
in the new variables $(t,y,K)$, and
approximate the corresponding flow to second order around the fixed point
$(t,y,K)=(0,0,0)$. We obtain
\begin{eqnarray}
\frac{\rmd t}{\rmd l}&= & 8K(1+t) - y^2\bigg(B^{(2)} - \frac{B^{(4)}}{16}\bigg) ,
\label{newflujo1} \\
\frac{\rmd y}{\rmd l}&= & -t y , \label{newflujo2} \\
\frac{\rmd K}{\rmd l}&= & -2 K - \frac{y^2}{64} B^{(4)} , \label{newflujo3}
\end{eqnarray}
where we have termed $B^{(n)} \equiv \tilde{B}^{(n)}(1,0)$. Introducing
new variables in order to bring flow (\ref{newflujo1})-(\ref{newflujo3})
into normal form
\cite{nayfeh}
\begin{eqnarray}
t &= & u_1 + \frac{1}{2} u_1^2 , \nonumber \\
y &= & u_2 , \nonumber \\
K &= & u_3 - \frac{B^{(4)}}{128} u_2^2 , \nonumber
\end{eqnarray}
we finally obtain
\begin{eqnarray}
\frac{\rmd u_1}{\rmd l}&= & 8 u_3  - B^{(2)} u_2^2 , \label{finflujo1} \\
\frac{\rmd u_2}{\rmd l}&= & -u_1 u_2 , \label{finflujo2} \\
\frac{\rmd u_3}{\rmd l}&= & -2 u_3 . \label{finflujo3}
\end{eqnarray}
The interesting feature of the approximate flow
(\ref{finflujo1})-(\ref{finflujo3}) is that
the third equation can be readily solved for the surface diffusion-like
variable $u_3$, as $u_3(l) = u_3(0) \exp(-2 l)$, thus making apparent
one of the qualitative features of the original RG flow, namely the fact that
surface diffusion is an irrelevant variable that decouples
(exponentially) fast from the flow for the lattice potential and
the surface tension. On the other hand, by taking the ratio between
the first and second equations in (\ref{finflujo1})-(\ref{finflujo3})
and integrating in $l$, one can check that the relation
\begin{equation}
u_1^2(l) = B^{(2)} u_2^2(l) - 16 \int_{l}^{\infty} u_1(l') u_3(l') {\rm d} l'
\label{sepa}
\end{equation}
defines a separatrix for the flow (\ref{finflujo1})-(\ref{finflujo3}),
generalizing the
well-known asymptotes of the sG hyperbolae \cite{ngreview} [see also panel (a) of
our Fig.\ \ref{fig:2}]. On e.g.\ the first quadrant of
the $(u_1,u_2)$ plane, flow lines below the separatrix flow onto the $u_2=0$
line of fixed points (irrelevant lattice potential, high temperature
behavior), whereas
flow lines above the separatrix flow towards large values of the
lattice potential $u_2$ (low temperature, massive phase). Thus, the separatrix
marks the temperature driven phase transition. Unfortunately, we have not
been able to produce a useful simpler expression for locus (\ref{sepa}),
not even within perturbation theory.

\begin{widetext}

\begin{figure}
\begin{center}
\includegraphics[width = 0.9\textwidth]{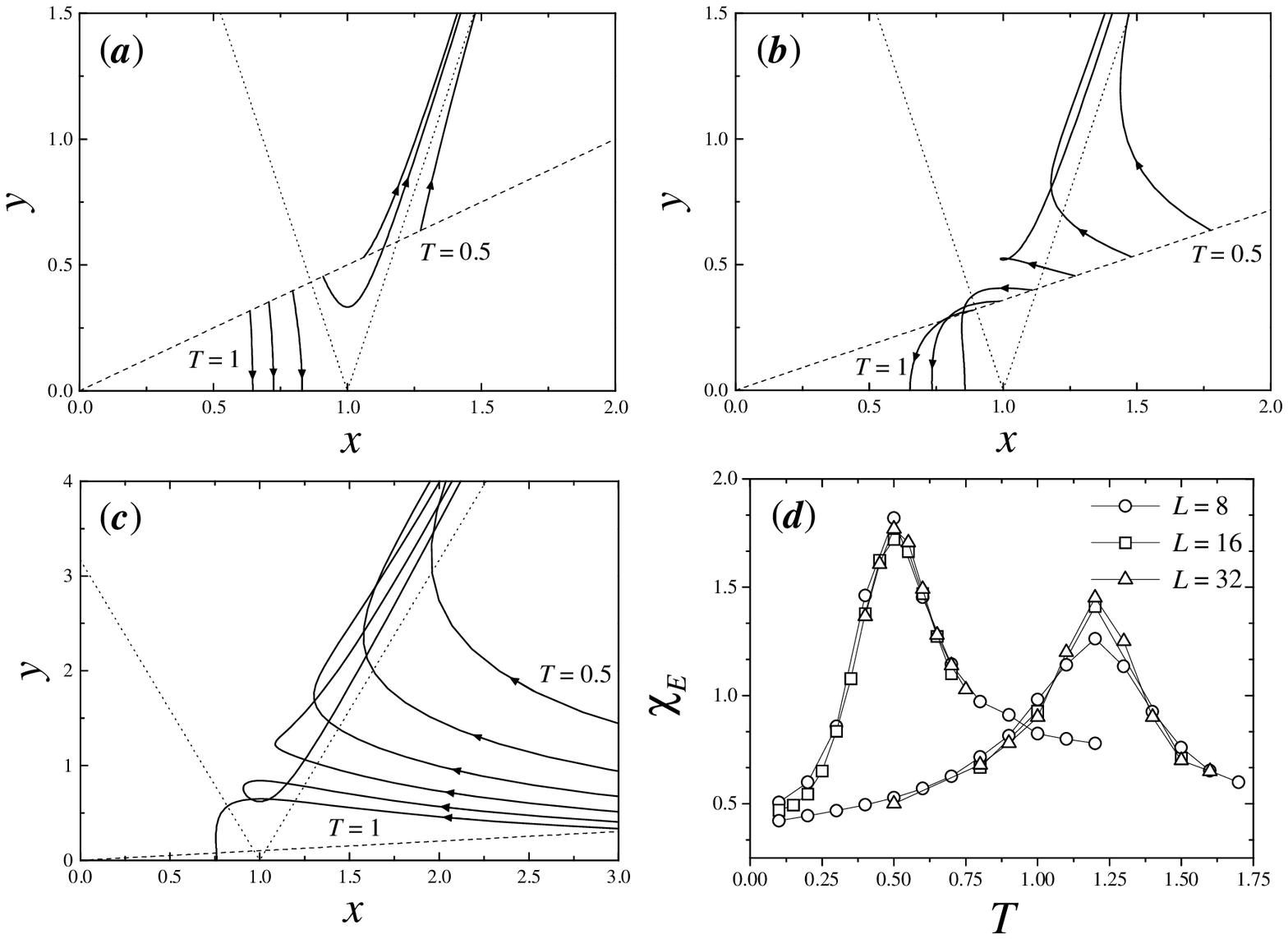}
\end{center}
\caption{Panels (a)-(c): $(x,y)$ projection of the RG flow
(\ref{finalflujodin1})-(\ref{finalflujodin3}) for $d=2$, using $\nu(0) =
V(0) = 1 = a_{\perp} = \Lambda= 1$, and (a) $\kappa(0) = 0$,
(b) $\kappa(0) = 10^{-2}$, (c) $\kappa(0) = 10^{-1}$. In panels (a)-(c),
initial conditions lie on the dashed line and the dotted line is the
separatrix for the pure sG flow. Solid lines correspond to
$T=$ 0.5, 0.6, 0.7, 0.8, 0.9, and $T=1$
right to left in (a) and (b), and top to bottom in (c). Panel (d): Specific
heat for the sG model (left curves) and for model (\ref{eqLangevinGR}) with
$\mu= \nu = 1$ and $\kappa = 0.5$ (right curves) for different system sizes.
All units employed are arbitrary.}
\label{fig:2}
\end{figure}

\begin{figure}
\begin{center}
\includegraphics[width = 0.5\textwidth]{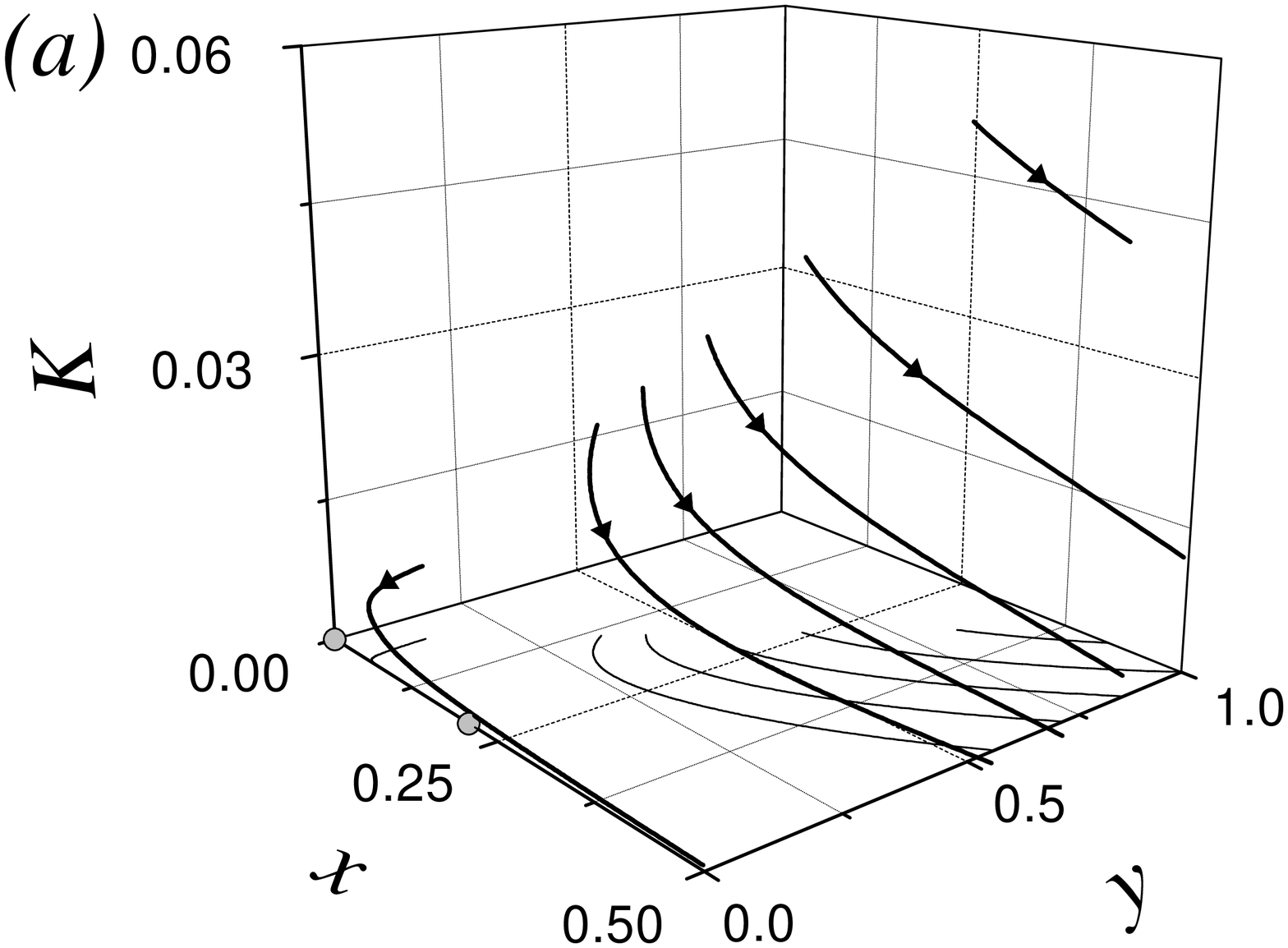}
\includegraphics[width = 0.5\textwidth]{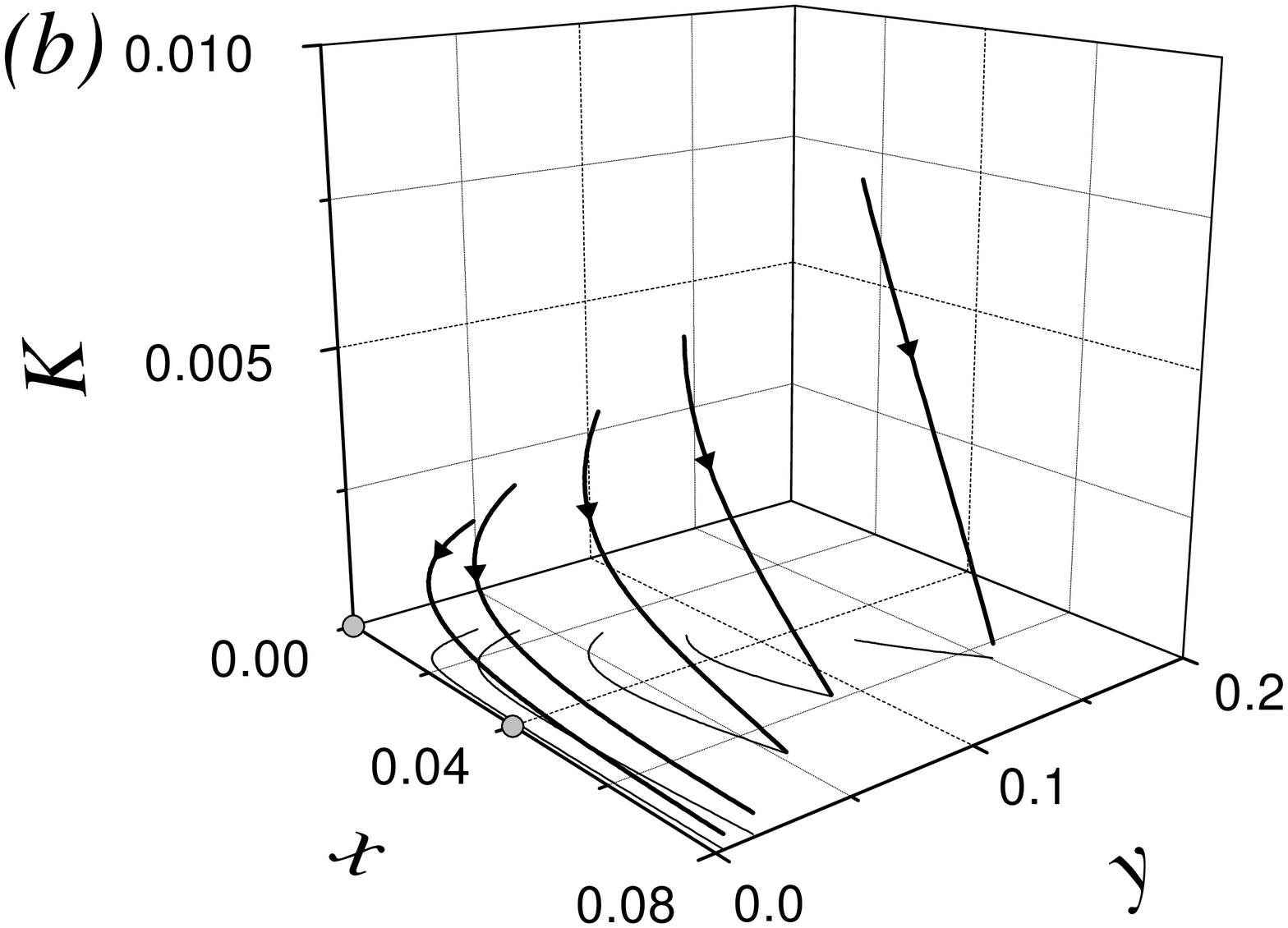}
\includegraphics[width = 0.5\textwidth]{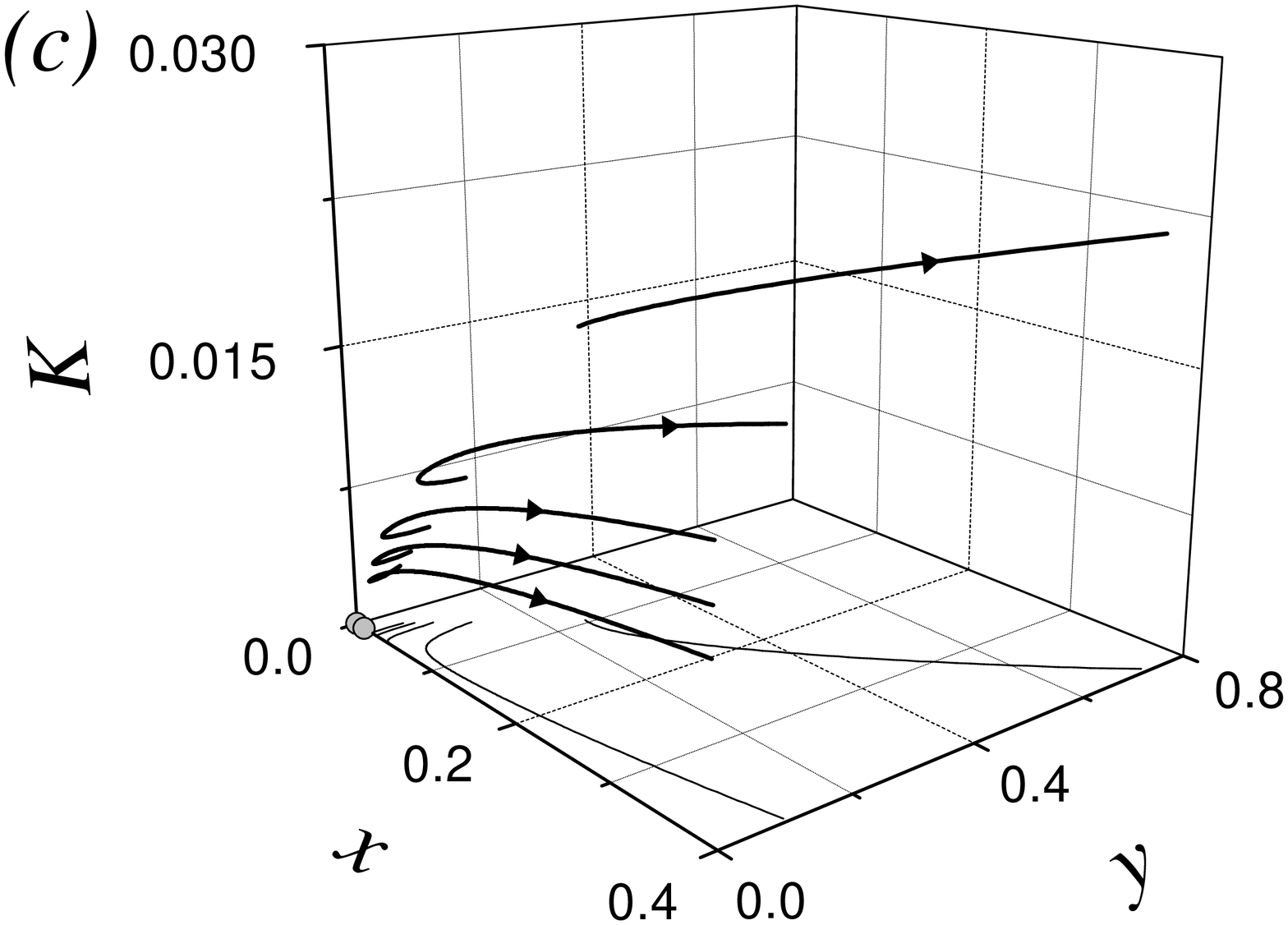}
\end{center}
\caption{RG trajectories from a numerical integration of
(\ref{finalflujodin1})-(\ref{finalflujodin3}) for $d=3$ (a), $d=4$ (b), and
$d=5$ (c) with $\nu(0)=0.5$, $\kappa(0)=0.25$, and $V(0)=0.1$.
Thick solid lines correspond, right to left, to
(a) $T=$ 1.5, 2, 2.5, 3, 3.5, 10; (b) $T=$ 10, 15, 20, 30, 40;
(c) $T=$ 5, 10, 15, 20, 25. Thin solid lines on the $(x,y)$ planes are
projections of the RG flow lines above them. Thick dots on the $x$ axis denote
both the origin and the point
$x = 2 {\cal S}_d \Lambda^{d-2}/[d (2\pi)^{d-1}]$ at which the rhs of the
flow equation for $y$ vanishes. Other parameters are as in Fig.\ \ref{fig:2}.
All units employed are arbitrary.}
\label{fig:2new}
\end{figure}

\begin{figure}
\begin{center}
\includegraphics[width = 0.5\textwidth]{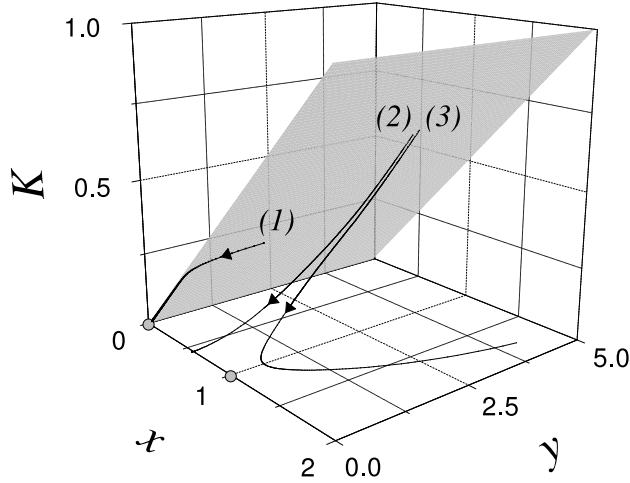}
\end{center}
\caption{RG trajectories from a numerical integration of
(\ref{finalflujodin1})-(\ref{finalflujodin3})
for the xMBE model [i.e.\ $\nu(0)=0$] with $L=128$, $\kappa(0)= 1$, and
$V(0)=0.1$.
Solid lines exemplify the three types of behavior depending on $T$:
(1) high $T$ phase, $l$MBE
scale invariant behavior ($T=10$);
(2) intermediate $T$ phase, EW scale invariant behavior
($T=0.46$);
(3) low $T$ massive phase ($T=0.45$).
For the sake of clarity, all three coordinates along line
(1) have been artificially expanded by a factor of 10.
The $x=2K$ plane of initial conditions appears shaded, and both the
origin and the $(1,0,0)$ point are signalled with thick dots.
Other parameters are as in Fig.\ \ref{fig:2}. All units employed are arbitrary.}
\label{fig:1}
\end{figure}

\begin{figure}
\begin{center}
\includegraphics[width = 0.7\textwidth]{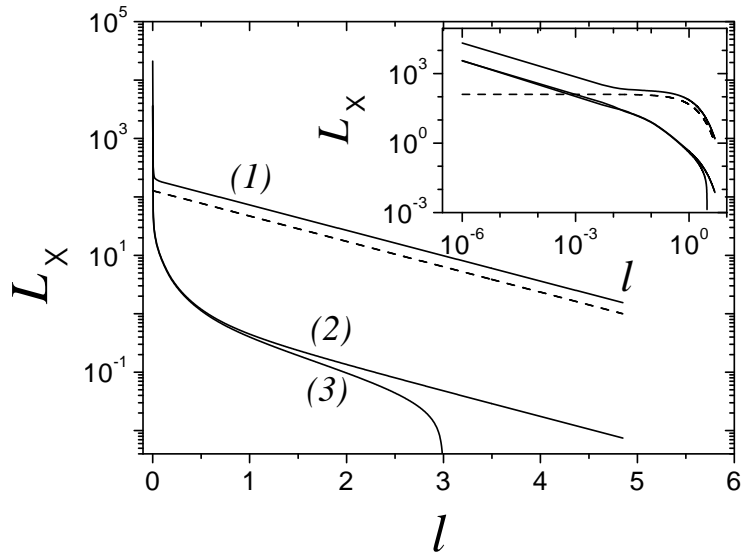}
\end{center}
\caption{Linear-log plot of the numerical RG flow of $L_{\times}(l)$ for
$d=2$, $L=128$. Other parameters are as in Fig.\ \ref{fig:1}. Solid lines
are for $T=$ 10, 0.46, and $T=0.45$, top to bottom.
For the sake of comparison, the
dashed line depicts the RG flow of $L_{\times}(l)$ for the linear system $V(0)=0$.
In all cases the flow is terminated for $l = \ln 128 \simeq 4.85$. All units
employed are arbitrary.
Inset: Blow-up of the same plot for small $l$ values, in log-log representation.}
\label{fig1'}
\end{figure}

\begin{figure}
\begin{center}
\includegraphics[width = 0.5\textwidth]{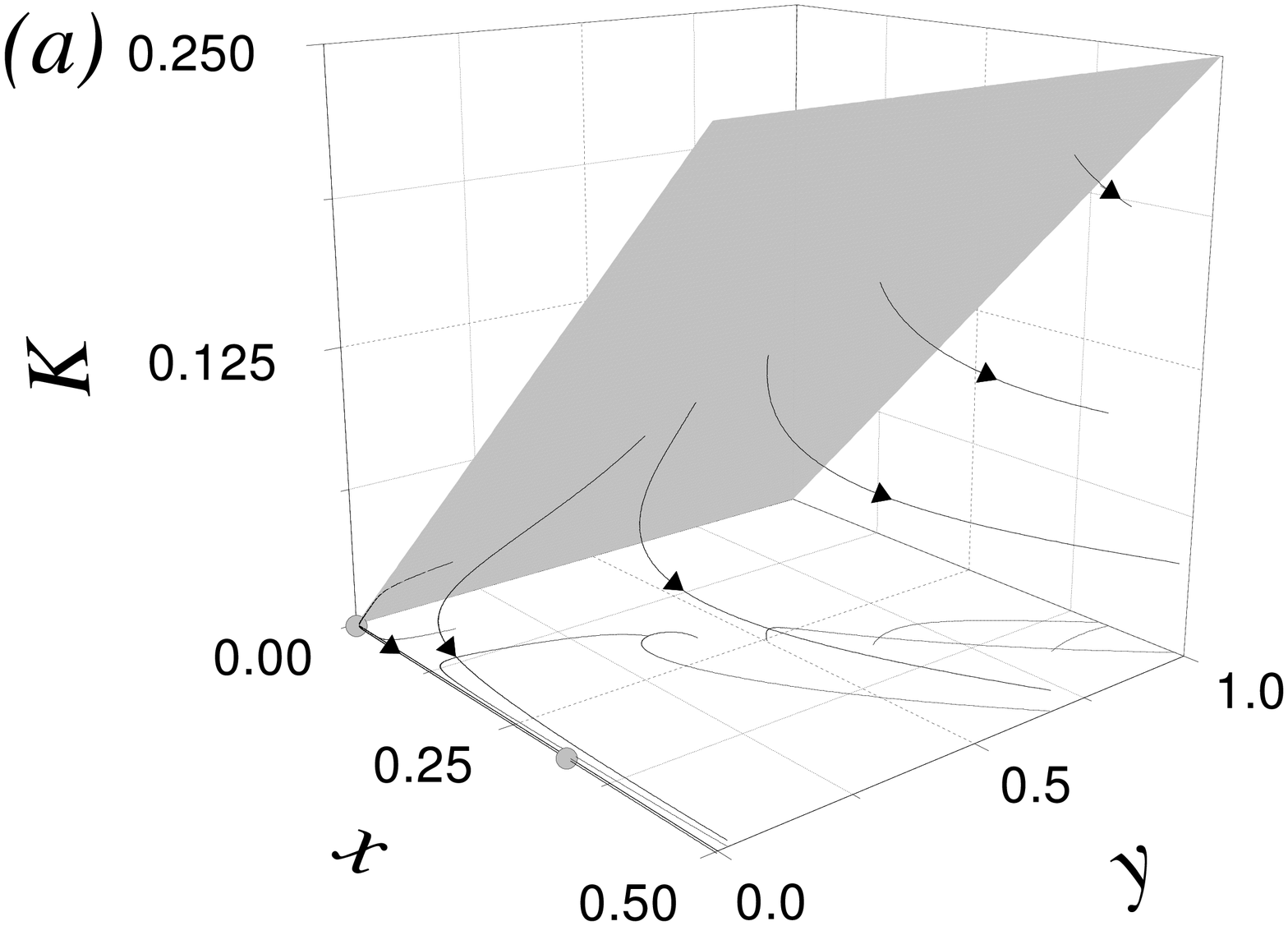}
\includegraphics[width = 0.5\textwidth]{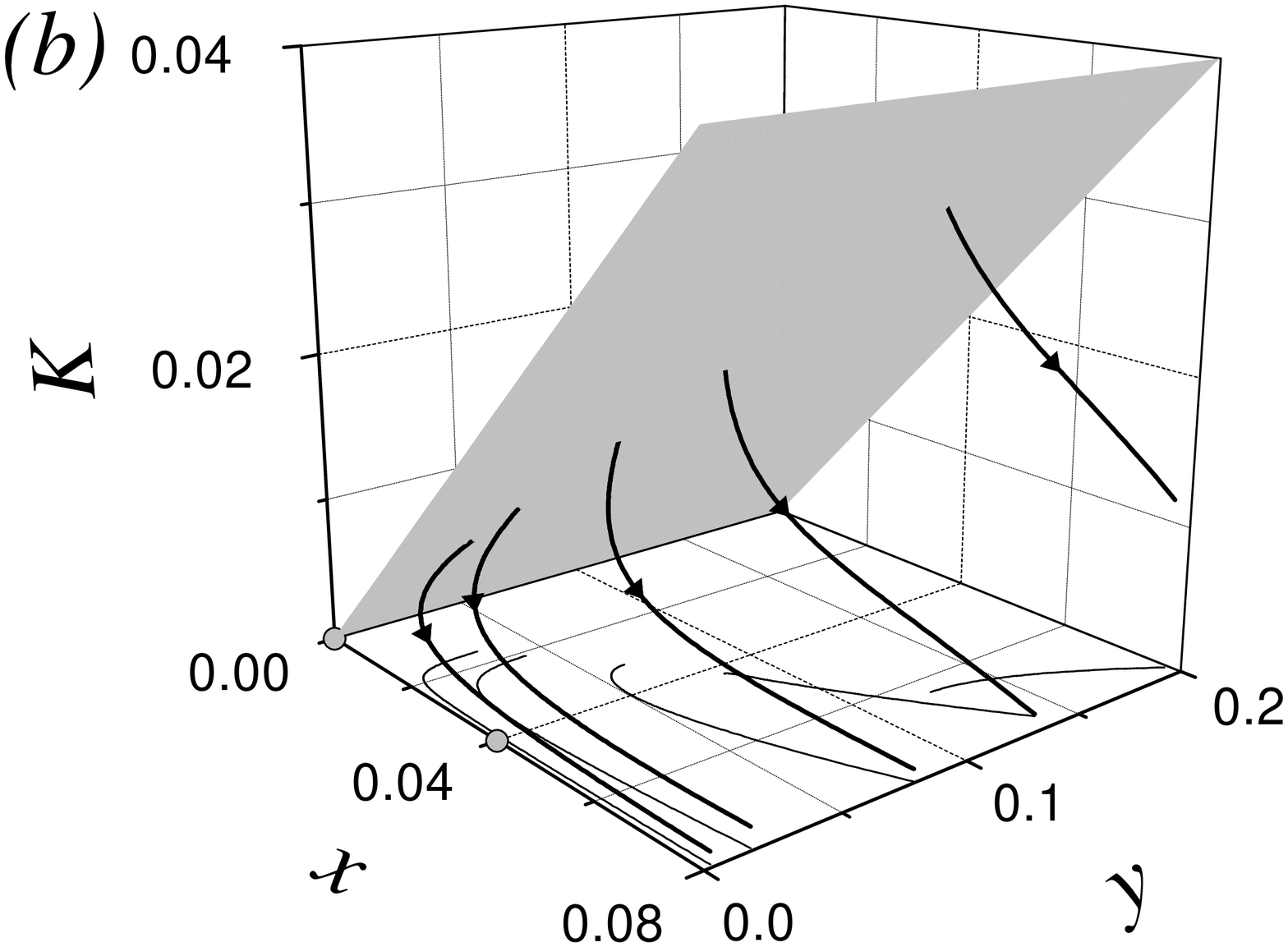}
\end{center}
\caption{RG trajectories from a numerical integration of
(\ref{finalflujodin1})-(\ref{finalflujodin3})
for the xMBE model [i.e.\ $\nu(0)=0$] with $\kappa(0)= 1$, and
$V(0)=0.1$. Panel (a) [(b)] corresponds to $d=3$ [$d=4$]. On each panel, solid
lines correspond to temperature values $T=$ 1.5, 2, 2.5, 3, 3.5, and $T=10$,
right to left. Thin solid lines on the $(x,y)$ plane are projections of the RG
flow lines above them.
As in Fig.\ \ref{fig:1}, the $x(0)=2K(0)$ plane of initial
conditions appears shaded. Thick dots on the $x$ axis denote both the
origin and the point
$x = 2 {\cal S}_d \Lambda^{d-2}/[d (2\pi)^{d-1}]$ at which the rhs of the
flow equation for $y$ vanishes. Other parameters are as in Fig.\ \ref{fig:2}.
All units employed are arbitrary.}
\label{fig:5new}
\end{figure}

\end{widetext}

\end{document}